\documentclass[aps,pra,twocolumn,amsmath,showpacs,preprintnumbers,superscriptaddress]{revtex4}
\usepackage{graphicx}
\usepackage{rotating}
\usepackage{amsmath}
\usepackage{color}
\usepackage{dcolumn}% Align table columns on decimal point
\usepackage{bm}% bold math

\begin{document}

\title{Collisions of Ultracold Trapped Cesium Feshbach Molecules\footnote{Dedicated to the memory of Prof.\ Vladilen S.\ Letokhov.}}
%==============================================================

%\title{Emergence of universal four-body states near an Efimov trimer}
%\title{Observation of a collisional suppression in ultracold halo dimers}
%\author{LevT Group}
%\altaffiliation[Also at ]{LENS and Dipartimento di Fisica, Universit\`{a} di
%Firenze, Firenze, Italy.}
\author{F. Ferlaino}
\author{S. Knoop}
\author{M. Berninger}
%\author{H. Sch\"obel}
\author{M. Mark}
\affiliation{Institut f\"ur Experimentalphysik and Zentrum f\"ur Quantenphysik, Universit\"at
 Innsbruck, % Technikerstra{\ss}e 25,
 6020 Innsbruck, Austria}
\author{H.-C. N\"{a}gerl}
\affiliation{Institut f\"ur Experimentalphysik and Zentrum f\"ur Quantenphysik, Universit\"at
 Innsbruck, % Technikerstra{\ss}e 25,
 6020 Innsbruck, Austria}
\author{R. Grimm}
\affiliation{Institut f\"ur Experimentalphysik and Zentrum f\"ur Quantenphysik, Universit\"at
 Innsbruck, % Technikerstra{\ss}e 25,
 6020 Innsbruck, Austria}
\affiliation{Institut f\"ur Quantenoptik und Quanteninformation,
 \"Osterreichische Akademie der Wissenschaften, 6020 Innsbruck,
 Austria}

\date{\today}

\begin{abstract}
We study collisions in an optically trapped, pure sample of ultracold Cs$_2$ molecules in various internal states. The molecular gas is created by Feshbach association from a near-degenerate atomic gas, with adjustable temperatures in the nanokelvin range. We identify several narrow loss resonances, which point to the coupling to more complex molecular states and may be interpreted as Feshbach resonances in dimer-dimer interactions. Moreover, in some molecular states we observe a surprising temperature dependence in collisional loss. This shows that the situation cannot be understood in terms of the usual simple threshold behavior for inelastic two-body collisions. We interpret this observation as further evidence for a more complex molecular structure beyond the well-understood dimer physics.
\end{abstract}

\pacs{33.20.-t, 34.50.Cx, 37.10.Pq}
%\keywords{}

\maketitle


%==============================================================
\section{Introduction}\label{intro}
%=============================================================================

Laser control of atoms and molecules has numerous important applications in physics \cite{Letokhov2007book}. A major research field emerged from the fascinating possibilities to cool and trap atoms by laser light \cite{Minogin1987book, Kazantsev1990book, Chu1998nlt, Cohentannoudji1998nlm, Phillips1998nll}. V.\ S.\ Letokhov gave very important contributions to the field; here the idea of confining atoms in standing light waves in 1968 \cite{Letokhov1968nod} and the first demonstration of atomic-beam cooling in 1981 \cite{Andreev1981rsa} represent two early examples for work that was well ahead of its time. More examples for the pioneering work of the laser cooling group at the Institute of Spectroscopy in Troitsk can be found in Refs.~\cite{Balykin1985rco, Balykin1988qss, Balykin1988coa, Grimm1990ooa}.

The great advances in the field of cooling and trapping in the 1980's and early 1990's \cite{Arimondo1992book} led to the attainment of Bose-Einstein condensation (BEC) in 1995 \cite{Anderson1995oob, Bradley1995eob, Davis1995bec} and the production of a degenerate Fermi gases in 1999 \cite{Demarco1999oof}. These achievements heralded the advent of a new era in physics, where ultracold atomic systems serve as well-controllable model systems to investigate a wide range of intriguing quantum phenomena \cite{Lewenstein2007uag, Bloch2007mbp, Giorgini2008tou, Chin2008fri}.

Molecular quantum gases emerged in the last decade. Great breakthroughs were achieved in the years 2002 and 2003, when molecular quantum gases could be produced in various systems of bosons \cite{Donley2002amc, Herbig2003poa, Durr2004oom, Xu2003foq} and fermions \cite{Regal2003cum, Strecker2003coa, Cubizolles2003pol, Jochim2003pgo}. The key to this success was the method of Feshbach association, where molecules are formed from colliding atom pairs by tuning a magnetic bias field across a Feshbach resonance \cite{Kohler2006poc, Chin2008fri, Krems2009book}. Molecular BEC was achieved at the end of 2003 \cite{Jochim2003bec, Greiner2003eoa, Zwierlein2003oob}. In the last few years great progress has been made in manipulating ultracold molecules, and they have found numerous intriguing applications \cite{Krems2009book}. Recent highlights are the transfer of Feshbach molecules to very deeply bound states \cite{Danzl2008qgo} and the creation of rovibrational ground-state molecules \cite{Ni2008ahp, Lang2008utm}.

With the experimental availability of ultracold trapped samples of molecules, it has become very important to understand their interaction properties.
%both from a fundamental and practical point of view.
For Feshbach molecules, the interactions are usually dominated by inelastic processes, as the molecules in a high rovibrational state carry a large amount of internal energy. Here the stability of molecules made of fermionic atoms is a remarkable exception \cite{Petrov2005spo, Giorgini2008tou}. Inelastic collisions between ultracold Feshbach molecules have been studied in various systems \cite{Mukaiyama2004dad, Chin2005oof, Syassen2006cdo, Knoop2008mfm, Ferlaino2008cbt, Inada2008cpo}. Most of these experiments did not show any significant dependence on the applied magnetic bias field. Two experiments on Cs$_2$, however, revealed novel magnetic-field dependent phenomena.  In Ref.~\cite{Chin2005oof} we observed loss resonances in dimer-dimer scattering, and in Ref.~\cite{Ferlaino2008cbt} we found a suppression of loss for weakly bound dimers in a halo state.

%----summary/inhalt----
In this Article, we study the collisional interactions between Cs$_2$ Feshbach molecules in more detail, building on our previous observations in Refs.~\cite{Chin2005oof, Ferlaino2008cbt}. Our new results provide further evidence for a more complex molecular structure beyond the well-understood dimer physics. In Sec.\,\ref{sec_overview}, we start with giving background information on Cs$_2$ Feshbach molecules.
In Sec.\,\ref{sec_res}, we present our results on collisional resonances in inelastic dimer-dimer scattering, which show up in different molecular states. In Sec.\,\ref{sec_T}, we report on the observations of a surprising temperature dependence of dimer-dimer collisions, which we find in some, but not all of the dimer states. In Sec.\,\ref{sec_outlook}, we finally discuss our results in view of more complex ultracold molecules.

\begin{figure*}
\begin{center}
\includegraphics[width=5.3in]{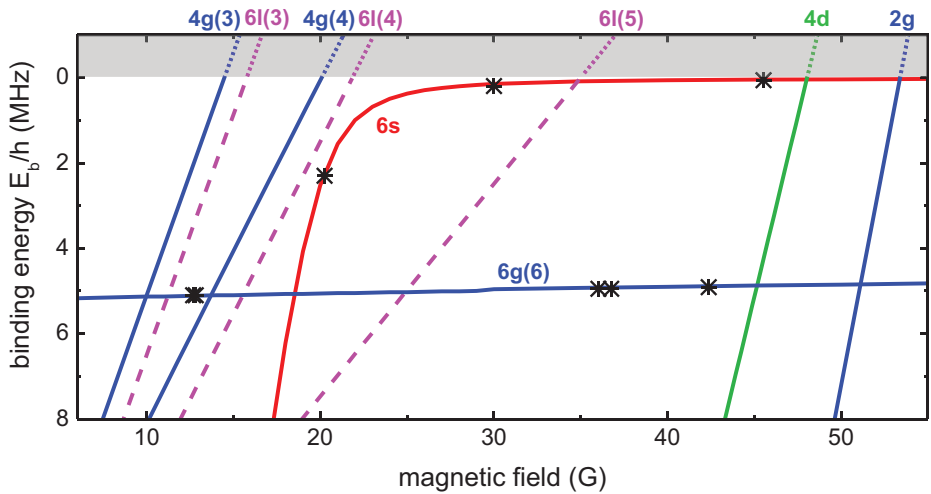}
\caption{Energy spectrum of weakly bound Cs$_2$ dimers versus magnetic field. The molecular states are labeled by the quantum numbers $f \ell (m_f)$ as discussed in Sec.~\ref{ssec_molstructure}; the quantum number $m_{f}$ is omitted for states with $m_{f}=f$ and $m_{\ell}=\ell$. The solid
lines represent $s$-, $d$-, and $g$-wave states. The intersections of the $d$- and $g$-wave states cause narrow Feshbach resonances, which can be used
for molecule production. The dashed lines represent $l$-wave states; they do not couple to the zero-energy threshold and therefore do not lead to Feshbach resonances. The stars mark the states and positions where we find the  narrow collisional resonances discussed in Sec.~\ref{sec_res}.
\label{fig1}}
\end{center}
\end{figure*}

%=============================================================================
\section{Ultracold Feshbach molecules made of cesium atoms}\label{sec_overview}
%=============================================================================

\subsection{Cesium as a quantum gas: brief history}

Cesium, the heaviest stable alkali species, was considered a candidate for Bose-Einstein condensation (BEC) already in the early 1990's \cite{Monroe1993moc}. After the attainment of BEC in Rb and Na \cite{Anderson1995oob,Davis1995bec}, it turned out that Cs atoms are subject to very special collisional interactions, which lead to anomalously fast spin flips. This inhibited all attempts to reach BEC by the standard magnetic trapping approach \cite{Soding1998gsr, Gueryodelin1998ibe, Arlt1998soc}. The unusual interactions were understood as a combination of resonant scattering with a very large indirect spin-spin coupling \cite{Leo2000cpo}.

It took until the year 2002 that BEC of Cs was reached in an optical dipole trap \cite{Weber2003bec}. In such a trap \cite{Grimm2000odt}, the atoms can be stored in the lowest internal state, where they are immune against two-body inelastic decay. It was crucial for the experiment to magnetically tune the $s$-wave scattering length in a way to optimize the collisional properties for the evaporative cooling process \cite{Kraemer2004opo, Kraemer2006efe}. Since then BEC of cesium has been realized in several other experiments \cite{Rychtarik2004tdb, Gustavsson2008coi, Hung2008aec}.

The successful production of near-degenerate samples or BECs of Cs opened up efficient ways to produce ultracold molecules via Feshbach association \cite{Herbig2003poa, Mark2005eco, Mark2007sou, Knoop2008mfm}. The basic principles and the many applications of this important association method are reviewed in \cite{Kohler2006poc, Krems2009book, Chin2008fri}.
Besides the creation of ultracold dimers, the remarkable scattering properties of Cs have proven very advantageous for the exploration of universal few-body phenomena related to Efimov states \cite{Kraemer2006efe, Ferlaino2008cbt, Knoop2009ooa, Ferlaino2009efu}.

\subsection{Energy structure of C$\textbf{\textrm{s}}$$_2$ Feshbach molecules}
\label{ssec_molstructure}

The internal structure of Cs$_2$ Feshbach dimers is particularly rich as compared to other alkali systems, and contributions by higher partial waves play an important role. Figure\,\ref{fig1} gives an overview of the molecular states relevant to the present work, covering the magnetic field region up to 55\,G and binding energies up to $h\times 8$\,MHz, where $h$ is Planck's constant. Zero energy corresponds to the dissociation threshold into two Cs atoms in the absolute hyperfine ground state sublevel $\vert F\!=\!3, m_{F}\!=\!3\rangle$.

%%%% Quantum numbers
Our notation $f \ell (m_f)$ for the angular momentum of the molecular states is explained in detail in Refs.~\cite{Chin2004pfs} and \cite{Mark2007sou}. The rotational angular momentum and its orientation are denoted by the quantum numbers $\ell$ and $m_{\ell}$, respectively; we follow the convention of labeling states with $\ell=0,2,4,6,8,\ldots$ as $s,d,g,i,l,\ldots$-wave states\,\cite{Russell1929ron}. The quantum numbers $f$ and $m_f$ refer to the internal spin of the molecular and its orientation, respectively.
All molecular states relevant to the present work obey $m_{f}+m_{\ell}=6$, which allows to simplify the notation by omitting one of these numbers. For states with $m_{f}=f$ and $m_{\ell}=\ell$, we use the notation $f \ell$ for brevity.

Feshbach resonances in general arise from the intersection of a molecular state with the atomic threshold \cite{Chin2008fri}. For Cs the observation of such resonances together with an elaborate quantum scattering model provided the essential information on the near-threshold molecular structure \cite{Chin2004pfs}. Higher-order interactions are particularly important for Cs \cite{Leo2000cpo} and lead to significant couplings between states up to $\Delta \ell = 4$. Therefore states up to $g$-wave character can couple to the $s$-wave scattering continuum, leading to observable resonances. The solid lines in Fig.~\ref{fig1} represent states up to $g$ waves ($\ell \le 4$), which were identified in this way \cite{Chin2004pfs}.
The dashed lines in Fig.\,\ref{fig1} represent $l$-wave
states ($\ell=8$), which do not cause observable Feshbach resonances but which can be populated via avoided crossings with $g$-wave states \cite{Mark2007siw, Mark2007sou, Knoop2008mfm}.

Coupling between molecular states with the same $f$ and $\ell$ in
general leads to very broad avoided crossings between molecular
states. The pronounced curvature of the $6s$ state in
Fig.\,\ref{fig1} is a result of such a strong crossing. In this
case, a weakly bound 6$s$-state with $F_{1}\!=\!3$ and $F_{2}\!=\!3$
couples to a 6$s$-state with $F_{1}\!=\!4$ and
$F_{2}\!=\!4$. Narrow avoided crossings arise when molecular
states of different $f$ and $\ell$ intersect \cite{Hutson2008acb}. Such crossings are not shown in Fig.\,\ref{fig1}, where the molecular states just intersect. Nevertheless, the existence of these weakly avoided crossings between molecular states of different $f\ell$ is crucial for a controlled state transfer by elaborate magnetic field ramps, as described in detail in Ref.~\cite{Mark2007sou}.

\subsection{Tunable halo dimers} \label{ssec_halodimers}

The state 6$s$ provides experimental access to the `halo' regime \cite{Jensen2004sar}, which is of particular interest in view of the universal properties of few-body systems \cite{Braaten2006uif}. For a halo dimer the only relevant length scale is given by the scattering length $a$, which describes the $s$-wave interaction between its two constituents. The size of the halo dimer is directly related to $a$ and the binding energy is $E_b = \hbar^2/(2 \mu a^2)$, where $\mu$ is the reduced mass. For ultracold gases, the characteristic interaction range is determined by the van der Waals potential and the halo regime is realized for binding energies well below an energy $E_{\text{vdW}}$ \cite{Kohler2006poc, Chin2008fri}; for the $s$-wave scattering length this condition is equivalent to $a \gg r_\text{vdW}$, where $r_\text{vdW}$ a characteristic length. For Cs, $E_{\text{vdW}} \approx h\times 2.7$~MHz and $r_\text{vdW} \approx 100\,$a$_0$, where $a_0$ is Bohr's radius.

In the 6$s$ state of Cs$_2$ we can conveniently control the binding energy via the magnetic field. The \textit{tunable halo dimers} realized in this way \cite{Mark2007sou, Ferlaino2008cbt} are unique probes to explore universal quantum phenomena \cite{Braaten2006uif}. As a recent example, we have observed a scattering resonance in the interaction of tunable halo dimers with free atoms \cite{Knoop2009ooa}, which we interpret as an atom-dimer resonance arising from the coupling to an Efimov trimer state \cite{Efimov1970ela, Kraemer2006efe}.

A binary collision between two halo dimers represents an \textit{elementary four-body process}. Collisional studies on halo dimers thus probe universal four-body physics, as has been demonstrated in the special case of halo dimers made of fermionic atoms in different spin states \cite{Petrov2004wbm, Inguscio2006ufg}. As a first step to explore universal four-body physics of identical bosons, we have recently measured collisional decay in a trapped sample of Cs$_2$ Feshbach molecules in the 6$s$ state \cite{Ferlaino2008cbt}. The experiments revealed a broad loss minimum around 30\,G ($a \approx 500 a_0$), for which  Ref.~\cite{Dincao2009ufb} presents a possible explanation in terms of universal four-body physics.

The full understanding of four-body systems is a present frontier of universal physics. Theoretical work has predicted pairs of four-body states tied to Efimov trimer states \cite{Hammer2007upo, Vonstecher2008fbl}. Very recently we could confirm the existence of these states by measured four-body recombination in an atomic cesium gas \cite{Ferlaino2009efu}. Studies on dimer-dimer interactions are of great interest for future experiments, as they hold great potential for future investigations, such as the observation of the resonant coupling of colliding dimers to tetramer states and systems of a trimer plus a free atom \cite{Dincao2009ufb}.

\subsection{Main experimental procedures} \label{ssec_procedures}

Our experimental procedures to produce an optically trapped sample of Cs$_2$ Feshbach molecules involve several stages. The main steps can be divided into the preparation of a near-degenerate atomic sample, the Feshbach association of the dimers, the purification of the molecular sample by removing remaining atoms, and the state transfer to the desired molecular state for the present investigation. Here we just give a short account of the main steps; further details can be found in Ref.\,\cite{Mark2007sou}.

The cesium atoms are first captured from an atomic beam into a standard magneto-optical trap, followed by an optical molasses cooling phase. A degenerate Raman sideband cooling stage \cite{Treutlein2001hba} then further increases the phase-space density and polarizes the atoms into the hyperfine ground state sublevel $\vert F=3,m_{F}=3\rangle$. Then the atoms are loaded into a large-volume optical dipole trap, followed by collisional transfer into a much tighter trap \cite{Weber2003bec, Kraemer2004opo}. The tight trap serves for evaporative cooling; it is formed by two crossed 1064-nm laser beams with waists of about 250 $\mu$m and 36 $\mu$m. We stop the cooling just before the onset of Bose-Einstein condensation to avoid too high atomic densities.

We magnetically associate the ultracold trapped cesium atoms to dimers on
Feshbach resonances \cite{Herbig2003poa, Xu2003foq, Durr2004oom, Kohler2006poc}. For this purpose we use three different resonances, the two $g$-wave resonances at $B=19.8$\,G and $53.4$\,G and the $d$-wave resonance at $47.9$\,G. To prepare a maximum number of molecules in the trap, it is
necessary to separate atoms and molecules as fast as possible,
since atom-dimer collisions dramatically reduce the lifetime of
the molecular sample \cite{Mukaiyama2004dad}. We remove the atoms
from the dipole trap using a `blast' technique similar to the schemes of
Refs.~\cite{Xu2003foq, Thalhammer2006llf}; the atoms are pushed out of the trap by resonant light, while the effect on the molecules remains negligible.

The molecular temperature can be set in a range between $40$\,nK and $300$\,nK by adjusting the temperature of the initial atomic sample. This can be done by variations in the evaporative cooling process and of the trap parameters \cite{Ferlaino2008cbt}. The number of trapped molecules is typically around 5000.

Other molecular states than the ones that we can directly access
through the Feshbach association schemes can be populated by
controlled state transfer \cite{Mark2007siw, Mark2007sou, Lang2008ctm}.
The experimental key is the precise control of Landau-Zener tunneling at avoided crossings through elaborate magnetic field ramps. By means of the ramp speed we can choose whether a crossing is followed adiabatically (slow ramp) or jumped diabatically (fast ramp). In this way, as demonstrated in Ref.~\cite{Mark2007sou}, we can efficiently populate any of the states shown in Fig.~\ref{fig1}.

To finally detect the molecules, we apply the standard method \cite{Chin2008fri} to dissociate the molecules by a reverse Feshbach ramp and to image the resulting atom cloud.

%% FRANCESCA'S PART
%\input{letokhov_results}

%=============================================================================
\section{Resonances in dimer-dimer scattering}\label{sec_res}
%=============================================================================
\begin{figure}
  \includegraphics[width=8.5cm] {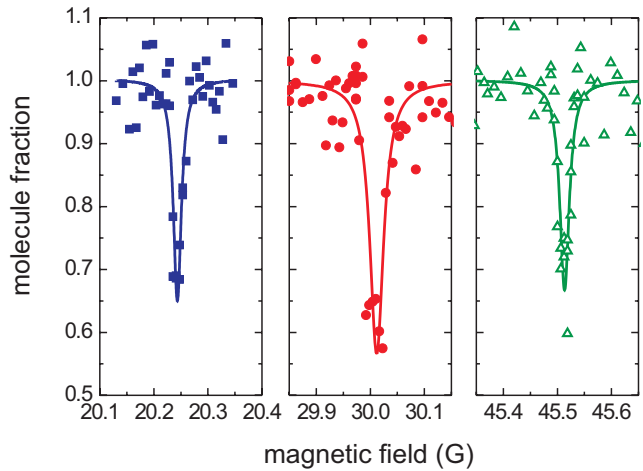}
  \caption{ Dimer-dimer resonances in the $6s$ state. The molecule fraction represents the number of $6s$ molecules after a 100-ms hold time in the dipole trap normalized to the corresponding average number detected off any resonance. The initial molecule number is typically 5000 at a temperature of about 150~nK. The loss resonances are fitted with Lorentzian profiles. The results of the fits are listed in Table \ref{tableresonances}.}
  \label{fig2}
\end{figure}

Studies of collisions between molecules can provide essential information on the interaction physics. We probe the system by studying collisionally induced loss in a trapped molecular sample. At ultralow temperatures, the dominant collisional mechanism  is relaxation into lower-lying bound states. Feshbach dimers carry a large amount of internal energy, which is in a relaxation event rapidly converted into kinetic energy . Since this released energy usually far exceeds the trap depth, both collisional partners can escape from the trap.
The resulting loss signal provides our experimental observable.

The relaxation rate depends on the possible decay channels, i.\,e.\,on the available lower-lying molecular state manifolds, and on the wave-function overlap between the initial and the final dimer state. For relaxation into deeply bound states, such a rate does not significantly depend on the applied bias magnetic field. This can be understood in terms of a simple molecular potential picture. The magnetic fields typically used in the experiments lead to energy shifts in the molecular potentials that are usually very small compared to the energy distance between the initial and final state. Consequently, the wave function overlap does not change substantially.
Typical relaxation rate constants are on the order of $10^{-10}$ - $10^{-11}$~cm$^{-3}/$s \cite{Mukaiyama2004dad, Chin2005oof, Syassen2006cdo, Knoop2008mfm, Inada2008cpo}.

Remarkable exceptions, in which the relaxation rate varies with the applied magnetic field, have been observed in collisions between halo molecules \cite{Jochim2003bec, Greiner2003eoa, Zwierlein2003oob, Ferlaino2008cbt}  and  when collisional resonances appear in dimer-dimer scattering \cite{Chin2005oof}. In the present Section, we  focus on the latter case.
\begin{figure}
\includegraphics[width=3.in]{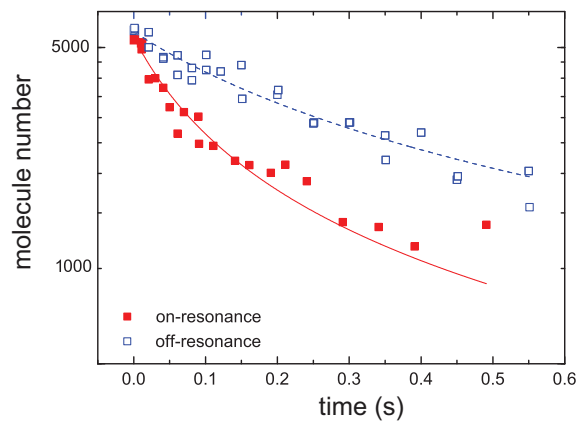}
\caption{\label{fig3}  Time evolution of the
number of $6s$ dimers at 20.24 G (on resonance,
squares) and at 20.11 G (off resonance, open squares). Fits
are based on the standard two-body loss equation; see text.}
\end{figure}

Collisional resonances in dimer-dimer scattering occur when the two dimers couple to  more complex bound states, involving either tetramer or trimer states.
When such a complex state approaches the threshold of two colliding dimers, either by crossing or merging, it induces a resonant enhancement of relaxation events and, consequently a resonant increase of losses. The number of dimers as a function of the magnetic field then shows  a sharp loss peak. In analogy to the case of coupling between two atoms  and a dimer state, these enhanced losses can be viewed as \emph{Feshbach-like resonances for ultracold molecules}. In a recent experiment, we observed such dimer-dimer resonances in a sample of  $g$-wave Cs$_2$ dimers \cite{Chin2005oof}.

Here, we  carry out a careful and systematic search of dimer-dimer resonances in a number of different molecular states by performing magnetic field scans with a typical step size of 5~mG. In particular, we explore the $6g(6)$ state in a magnetic field range between 10 to 13~G and 25 to 45~G and the $4g(4)$ state, the $4d$ state and the $6s$ halo  state up to $5$~MHz binding energy.

We measure the collisional resonances in dimer-dimer scattering by following  an experimental procedure similar to Ref.\,\cite{Chin2005oof}. We first produce a  sample of optically trapped dimers in a desired molecular states, as outlined in Sec.\,\ref{ssec_procedures}. We then hold the dimers in the dipole trap for a certain time, typically 100 to 200 ms. Finally, we record the number of remaining dimers by standard dissociation detection; see Sec.\,\ref{ssec_procedures}. The measurement is then repeated for various magnetic field values in the range of interest and for various molecular states.

Figure \ref{fig2}  shows the three dimer-dimer resonances found in the $6s$ state.
The observed loss peaks are typically very pronounced, very narrow, and symmetric.
By fitting the loss peaks  with  Lorentzian profiles, we precisely determine  the positions and we extract the widths of the resonances, as listed in Table  \ref{tableresonances}.
We use the Breit-Rabi formula to determine the magnetic-field value from a
measurement of the resonant frequency corresponding to the $|F\!=\!3,
m_{F}\!=\!3\rangle \rightarrow |F\!=\!4,m_F\!=\!4\rangle$ atomic hyperfine
transition.  Differently from Ref.\,\cite{Chin2005oof}, in which a levitating magnetic field was employed to support the dimers against gravity, we here use  a purely optical trap. The present setup does not suffer from the corresponding inhomogeneity and thus improves the magnetic field resolution to about 10 mG, compared to 150~mG resolution obtained with the levitation field.

The enhancement of losses can be characterized by studying the time dependence
of the molecular decay in the optical trap. Figure \ref{fig3} shows the decay curve for $6s$ dimers at the 20-G resonance (squares) and slightly below resonance (open squares).
Both on and off resonance, we observe a non-exponential decay of the dimer number. The decay of the trapped dimer sample is   well described by the usual two-body rate equation
\begin{equation}
\dot{N}=-\alpha_{\text{rel}}\bar{n} N,
\label{Eq1}
\end{equation}
where $N$ indicates the number of dimers and $\alpha_{\text{rel}}$ the relaxation rate coefficient. The mean molecular density $\bar{n}$ is given by $\bar{n}=\left[m \bar{\omega}^2/(2\pi
k_B T)\right]^{3/2}N$ with $m$ being the atomic mass and $\bar{\omega}$
denoting the geometric mean of the trap frequencies.
Typically, we observe the relaxation rate coefficient to increase by  about a factor  of 5 on resonance.

We perform similar magnetic field scans in the states $6g(6)$, $4d$, and $4g(4)$.
In the $6g(6)$ state, we both confirm the existence of the resonances  observed in previous experiments \cite{Chin2005oof}, and we identify  three additional new resonances in the magnetic field range between 25 to 45~G. Up to binding energies of 5~MHz, we do not observe any resonance for dimers in the $4d$ and $4g(4)$ state. Our results are listed in Table  \ref{tableresonances}. For illustrative purposes, we also mark the resonance positions in Fig.\,\ref{fig1} (stars).

An important open question is whether the observed resonances are related to tetramer or trimer states, coupling to the dimer-dimer threshold. The observation of  resonance shapes that are symmetric suggests a coupling to a tetramer state. In fact, one can expect that the coupling with a trimer plus a free atom would  lead to an asymmetric shape because of the continuum of energy \cite{Dincao2009ufb}. However, further investigations are needed to clearly distinguish between these two processes.

\begin{table}
\begin{ruledtabular}
\begin{tabular}{c|ccc}

dimer branch &  $B({\rm G})$ & $\Delta B({\rm mG})$& $a(a_0)$\\
\hline $6s$    & 20.24 (1)& 18 (3) & 179 \\
            &   30.01 (1) & 31 (4) & 552 \\
            &    45.52 (1)& 21 (4) & 893\\
\hline $6g(6)$    & 12.67 (1)\footnotemark[1]& 5 (2)& \\
            &   12.82 (1)\footnotemark[1]& 19 (2) &\\
            & 36.06 (1)& 10 (2) &\\
            &    36.80 (1)& 19 (5) &\\
            &    42.43 (1)&  9 (2)&\\
\end{tabular}
\end{ruledtabular}
\footnotetext[1]{The two resonance positions in the $6g(6)$ state near 13~G slightly differ from the values of Ref.\,\cite{Chin2005oof}. In this previous experiment, the levitation field introduced an uncontrolled offset.}
\caption{\label{tableresonances} Observed dimer-dimer resonances. The resonances are found in the states $6s$ and $6g(6)$, while no resonances are found in the states $4d$ and $4g(4)$.  The locations and the widths of the loss signals, together with their statistical errors, are obtained from Lorentzian fits. For the $6s$ state, the Table also gives the corresponding value of the scattering length \cite{Lange2009doa}.}
\end{table}

\begin{figure*}
\begin{center}
\includegraphics[width=4.8in]{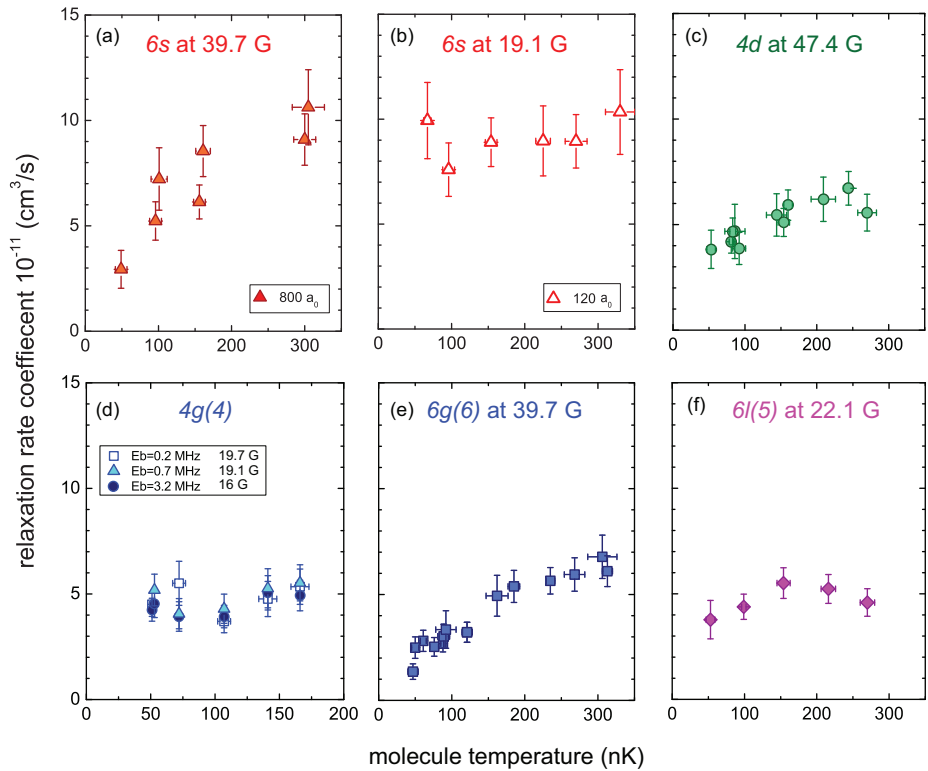}
\caption{Temperature dependence of the relaxation rate
 coefficient $\alpha_{\text{rel}}$ for different molecular states: the $6s$ state in the halo (a) and non-halo regime (b), the $4d$ state (c), the $4g(4)$ state (d) for different binding energies, the $6g(6)$ state (e), and the $6l(5)$ state (f). The data in (a) and (b) are from \cite{Ferlaino2008cbt}. \label{fig5}}
\end{center}
\end{figure*}

%=============================================================================
\section{Temperature dependence of collisional loss}\label{sec_T}
%=============================================================================

For inelastic processes in the ultracold regime, the relaxation rate coefficient is usually independent of the  collision energy, or the temperature $T$ of the sample. This threshold law applies for $s$-wave collisions and for $k_{B}T$ smaller than  all other energy scales in the system \cite{Weiner1999eat}.  The essential point behind this threshold law is  that relaxation processes into deeply bound states  release such a large amount of internal energy that  the collisional energy plays no role. This picture breaks down when another state is energetically close to the threshold of the two colliding particles.

In our previous experiments on collisions between $6s$ dimers, we found a surprising temperature dependence of the relaxation rate coefficient  \cite{Ferlaino2008cbt}. In the non-halo regime, $\alpha_{\text{rel}}$ showed the expected constant behavior, while a clear increase with temperature was observed  in the halo regime.  A recent theoretical work suggests that the temperature dependence observed in the halo state is related  to the existence of a universal trimer state, which lies energetically slightly above the zero-energy threshold of two colliding atoms \cite{Dincao2009ufb}.

We here raise  the question whether the unusual temperature behavior is a property unique to halo states or whether it can also occur for other Feshbach molecules. We thus probe the temperature dependence of the relaxation rate coefficient for different dimer states, investigating the states $6s$, $4d$, $4g(4)$, $6g(6)$, and $6l(5)$.

We measure the  time dependence of the molecular decay  for various temperatures, recording decay curves similar to Fig.\,\ref{fig3}.  The relaxation rate coefficient $\alpha_{\text{rel}}$ is extracted by fitting the decay curve with the usual two-body rate equation, Eq.\,(\ref{Eq1}).
Our findings are summarized in Fig.\,\ref{fig5}. We observe a clear temperature dependence in  two cases, while the other three cases follow the threshold law expectation of a constant rate coefficient.  In one case the result is ambiguous. The two cases with a clear temperature dependence are the $6s$ state at about 39.7~G (Fig.\,\ref{fig5}a) and the $6g(6)$ state (Fig.\,\ref{fig5}e); here the relaxation rate increases with temperature,  roughly following a $\sqrt{T}$-dependence.  Constant rate coefficients are observed  in the $6s$ state at  about 19.1~G, the $4g(4)$ state, and the $6l(5)$ state. The ambiguous case is the $4d$ state, which seems to show a weak temperature dependence; here the situation may be obscured by the proximity of the $d$-wave Feshbach resonance at 47.78~G \cite{Lange2009doa}.

Motivated by our previous observation that the temperature dependence of $\alpha_{\text{rel}}$ changes with the magnetic field in the $6s$ state \cite{Ferlaino2008cbt}, we perform similar investigations in the $g$-wave states. In $4g(4)$, we check three points between 16 and 19.7~G, finding essentially the same behavior of a constant rate coefficient; see Fig.\,\ref{fig5}d.
In $6g(6)$, we inspect the range between 25 and 45~G (data not shown), the temperature dependence being preserved over the full range.  Comparing the observations on the two $g$-wave states with the behavior of the $6s$ state, which in contrast to all other states has a curvature, we find an interesting systematics.
The temperature dependence shows up when the state is parallel to the atomic threshold, which means that threshold and molecular state have the same magnetic moment.  This is the case for the $6g(6)$ state and the $6s$ state in the halo regime. In contrast, we do not observe the temperature dependence when the magnetic moments are different, which is the case  for the $4g(4)$ state, the 6$l$(5) state, and for the $6s$ state in the non-halo regime. It is also interesting to note that the two states that show a temperature dependence are also the only two for which resonances in dimer-dimer scattering have been observed.  These observations may just be accidental coincidences, but they may also have a deeper physical reason.

%=============================================================================
\section{Conclusion and outlook}\label{sec_outlook}
%=============================================================================

We have reported on two phenomena where ultracold dimer-dimer collisions point to a more complex structure beyond the two-body physics of dimers. Understanding this structure is not only important from a fundamental point of view, it also holds potential for future experimental applications.

The occurrence of narrow loss resonances is most likely explained by coupling to tetramer states. The analogy of this situation to Feshbach resonances in atom-atom collisions \cite{Chin2008fri} points to potential applications, which can be found in controlling dimer-dimer interactions and in the association of more complex molecules. Because of the many open decay channels such molecules will be inherently unstable, but situations may exist where loss suppression mechanisms \cite{Petrov2004wbm,Dincao2008som} will enable observation times long enough for experimental applications. Then it may be possible to perform a magnetic field ramp to associate tetramers from colliding dimers. A related possibility would be a controlled rearrangement reaction where two dimers are converted into a trimer and a free atom \cite{Dincao2009ufb}. Further possible applications of collisional resonances arise from the prospect to use strong dissipation in dimer-dimer collisions for realizing strongly correlated states of matter \cite{Syassen2008sdi}.

The temperature dependence observed in some of the investigated collision channels provides hints on the existence of thresholds which are energetically slightly above the zero-energy threshold of two colliding dimers; this again points to the role of more complex molecular states. For the universal 6$s$ halo dimer state, the existence of a trimer state indeed provides a plausible explanation for the observed behavior \cite{Dincao2009ufb}. For the non-universal 6$g$(6) state, which shows a very similar temperature dependence over a wide magnetic field range, we can only speculate that a non-universal trimer or tetramer state may closely follow the dimer state's energy.

We are only beginning to understand the complex interaction properties of ultracold trapped molecules, but we already see intriguing phenomena with potential applications in this emerging research field.

%=============================================================================
\section{A personal note}
%=============================================================================

I (R.G.) had the great opportunity to work with Prof.\ V.\ S.\ Letokhov as a postdoctoral researcher at the Institute of Spectroscopy in Troitsk in 1989-1990, and then as guest scientist during several visits in the early 1990's. This was an extremely inspiring time for me, which laid the ground for my later work on optical dipole forces and atomic quantum gases. Prof.\ Letokhov was very much interested in the possibility to extend the efficient laser cooling methods to molecules. I remember him pointing out many times the great potential of physics with laser-manipulated molecules. I always had his words in mind when, many years later, I directed my scientific work towards molecules in ultracold quantum gases.

\begin{acknowledgments}
We thank C. Chin, S. D\"urr, and J. D'Incao for stimulating discussions. We acknowledge support by the Austrian Science Fund (FWF) within SFB 15. F.\,F.\ was supported within the Lise Meitner program of the FWF, and S.\,K.\ was supported by the European Commission with a Marie Curie Intra-European Fellowship.
\end{acknowledgments}

%\bibliographystyle{apsrev}

%\bibliography{../ultracold,fixultracold,letokhov_RG,letokhov_FF}

\end{document}